\documentclass{aa}
\usepackage{graphicx}                    
\usepackage{natbib}                     
\usepackage{times}
\usepackage{amsmath}
\usepackage{amssymb}                    
\usepackage{color}                       
\usepackage{url}                         

\usepackage{verbatim}

\defcitealias{Deheuvels_ea:2014}{D+14}
\defcitealias{Mosser_ea:2012}{M+12}

\begin{document}

\title{Angular momentum transport efficiency \\ in post-main sequence low-mass stars}
\titlerunning{Angular momentum transport efficiency in post-main sequence low-mass stars}

\author{F.~Spada$^1$, M.~Gellert$^1$, R. Arlt$^1$ \and  S.~Deheuvels$^{2,3}$ }
\authorrunning{F. Spada,  M. Gellert, R. Arlt \and S. Deheuvels }
       
%
\institute{
$^1$ Leibniz-Institut f\"ur Astrophysik Potsdam, An der Sternwarte 16, D-14482 Potsdam, Germany \\
\email{fspada@aip.de} \\
$^2$ Universit\'e de Toulouse, UPS-OMP, IRAP, 31028 Toulouse, France \\
$^3$ CNRS, IRAP, 14 avenue Edouard Belin, 31400 Toulouse, France  \\
}

\date{Received; accepted}
 
\abstract{ 
Using asteroseismic techniques, it has recently become possible to probe the internal rotation profile of low-mass ($\approx 1.1$--$1.5\, M_\odot$) subgiant and red giant stars.
Under the assumption of local angular momentum conservation, the core contraction and envelope expansion occurring at the end of the main sequence would result in a much larger internal differential rotation than observed. 
This suggests that angular momentum redistribution must be taking place in the interior of these stars. 
}{
We investigate the physical nature of the angular momentum redistribution mechanisms operating in stellar interiors by constraining the efficiency of post-main sequence rotational coupling.
}{
We model the rotational evolution of a $1.25\; M_\odot$ star using the Yale Rotational stellar Evolution Code.
Our models take into account the magnetic wind braking occurring at the surface of the star and the angular momentum transport in the interior, with an efficiency dependent on the degree of internal differential rotation.
}{
We find that models including a dependence of the angular momentum transport efficiency on the radial rotational shear reproduce very well the observations.
The best fit of the data is obtained with an angular momentum transport coefficient scaling with the ratio of the rotation rate of the radiative interior over that of the convective envelope of the star as a power law of exponent $\approx 3$.
This scaling is consistent with the predictions of recent numerical simulations of the Azimuthal Magneto-Rotational Instability.
}{
We show that an angular momentum transport process whose efficiency varies during the stellar evolution through a dependence on the level of internal differential rotation is required to explain the observed post-main sequence rotational evolution of low-mass stars.
}


\maketitle

\section{Introduction}
\label{intro}

Rotation is a very important property of stars, as it can significantly affect stellar structure and evolution in a variety of ways (e.g., directly, \citealt{Sills_ea:2000}; by enhancing element mixing, \citealt{Pinsonneault:1997}; by powering dynamo action and magnetic activity, \citealt{Noyes_ea:1984}). 

Solar-like stars show a significant rotational evolution during their pre-main sequence (PMS) and main sequence (MS) lifetime and beyond, as their magnetized stellar winds effectively drain angular momentum from their surfaces \citep{Schatzman:1962,Kraft:1967, Skumanich:1972}. 
Simple rotational evolution models can reproduce the basic features of this spin-down \citep[see, e.g.,][]{Gallet_Bouvier:2013}, when also taking into account the occurrence of structural readjustments (e.g., the gradual development of an inner radiative zone during PMS contraction), and of angular momentum transport within the interior, typically ensuring an efficient rotational coupling within a time scale of a few hundreds of Myr \citep[see also][]{MacGregor_Brenner:1991,Denissenkov_ea:2010,Spada_ea:2011}.

Although it has been possible to measure the surface rotation of stars for a long time \citep[see, e.g.,][and references therein]{Kraft:1970}, much less is known about the rotational state of stellar interiors.
Through helioseimology, the solar rotation profile has been mapped from the surface down to $\approx 0.2$ solar radii \citep{Schou_ea:1998,Howe:2009}, showing a remarkably low radial differential rotation; similarly, using asteroseismic techniques, it has recently become possible to place constraints on the radial differential rotation in the interior of other stars.
The analysis of the rotational splittings carried out on six solar-like stars by \citet{Nielsen_ea:2014} ruled out strong radial rotational gradients.
Similarly,  \citet{Benomar_ea:2015} reported ratios of interior to surface rotation rates smaller than a factor of $2$ in a sample of stars of mass between $1.1$ and $1.6\, M_\odot$, independently of their age.
Quantitative constraints on the degree of internal differential rotation can also be placed for stars sufficiently evolved to show mixed modes \citep{Dupret_ea:2009,Beck_ea:2012,Deheuvels_ea:2012,Mosser_ea:2012,Deheuvels_ea:2014}.

Observations thus strongly support an efficient angular momentum redistribution in the interior of low- and intermediate-mass stars, leading to a state of almost uniform rotation within the mature stages of their MS lifetime.
The physical nature of the processes responsible for this rotational coupling, however, remains elusive. 
Both internal gravity waves \citep[see][for a review]{Mathis:2013}, and magnetic fields \citep[e.g.,][]{Charbonneau_MacGregor:1993,Ruediger_Kitchatinov:1996} have been proposed as viable mechanisms to redistribute angular momentum in the interior of MS stars.

The rotational evolution beyond the TAMS is equally non-trivial.
As the stellar core contracts and the envelope expands at the end of the hydrogen burning phase, local angular momentum conservation would require the development of a strong differential rotation. 
The moderate differential rotation observed in KIC $7341231$ \citep{Deheuvels_ea:2012} or KIC $8366239$ \citep{Beck_ea:2012} is incompatible with this scenario, suggesting that efficient angular momentum transport is taking place during the post-main sequence (poMS) phase as well.
Purely hydrodynamical effects, such as meridional circulation and the shear instability, have been shown to be insufficient to reconcile the theoretical models with this observed behavior \citep{Eggenberger_ea:2012,Ceillier_ea:2013,Marques_ea:2013}. 
Models including angular momentum transport mediated by the so-called Tayler-Spruit Dynamo \citep{Cantiello_ea:2014} and by gravity waves \citep{Fuller_ea:2014} also produce too fast core rotation in subgiants in comparison with the observations.

The stars in the samples of \citet[hereafter \citetalias{Deheuvels_ea:2014}]{Deheuvels_ea:2014} and \citet[\citetalias{Mosser_ea:2012} in the following]{Mosser_ea:2012}, combined together, offer the unique opportunity to trace the evolution of core--envelope differential rotation from immediately after the TAMS to the giant branch stage.

In this work, we focus on modeling the poMS rotational evolution of stars of approximately solar mass ($M\lesssim 1.5\, M_\odot$), using the Yale Rotational stellar Evolution Code \citep[YREC;][]{Pinsonneault_ea:1989,Demarque_ea:2008}. 
In particular, motivated by the results of recent numerical simulations of the Azimuthal Magneto-Rotational Instability (AMRI; \citealt{Ruediger_ea:2015}, Gellert et al., in preparation), we wish to test the dependence of the angular momentum transport efficiency on the degree of internal differential rotation.

The AMRI is a destabilization of hydrodynamically stable differential rotation by current-free toroidal magnetic fields \citep{Ruediger_ea:2007}. 
In contrast to the magneto-rotational instability of an axial field \citep[see][]{veli_57}, it is naturally nonaxisymmetric. 
The instability extracts its energy from the differential rotation, thus working more effectively for steeper rotation profiles, and disappearing in the presence of solid-body rotation. 
The turbulent viscosity generated by the AMRI can be very effective at transporting angular momentum \citep{Ruediger_ea:2015}, with the magnetic contribution due to Maxwell stresses strongly dominating over the kinetic component. 
As a consequence, mixing is much less enhanced by the AMRI than angular momentum transport, and the Schmidt number (the ratio of turbulent viscosity and turbulent element diffusion coefficient) is large enough not to speed up stellar evolution significantly \cite{schatz_77,leb_87,brott_08}.
A distinctive property of the AMRI-induced viscosity is its dependence on the angular velocity shear present in the region where the instability develops.
As a consequence, the time scale for the quenching of the differential rotation is not constant in time, but increases as the rotation profile becomes flatter. 
This is a specific prediction that can be tested using our formulation. 

The paper is organized as follows: in Section \ref{obs} we introduce the data used to constrain our models; in Section \ref{models} we describe our models and the implementation of the turbulent angular momentum diffusion dependent on the internal differential rotation. 
Our results are presented in Section \ref{results} and discussed in Section \ref{discussion}.
We summarize our conclusions in Section \ref{conclusions}.

\begin{figure}
\begin{center}
\includegraphics[width=0.49\textwidth]{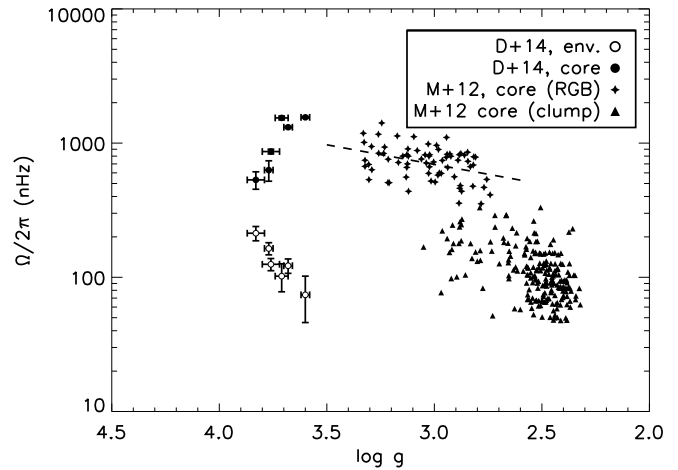}
\caption{Asteroseismic constraints on the internal differential rotation of evolved low mass stars stars used in this work. The broken line is a fit of the red giants in the \citetalias{Mosser_ea:2012} sample.}
\label{data}
\end{center}
\end{figure}

\section{Asteroseismic constraints on poMS rotational coupling} 
\label{obs}

We consider asteroseismic constraints on the poMS rotational evolution of low-mass stars from the following two studies (see Figure \ref{data}).

The \citetalias{Deheuvels_ea:2014} sample contains six stars of mass between $1.1$ and $1.5 \; M_\odot$ and approximately solar composition. 
For these stars, the surface gravity, $\log g$, and the core and envelope angular velocities, $\Omega_g$ and $\Omega_p$ (filled and empty circles, respectively), have been determined from individual asteroseismic modelling (here we loosely refer to the rotation rate averaged over the $g$-mode and the $p$-mode cavities, respectively, as ``core" and ``envelope" rotation rates; see section $6$ of \citetalias{Deheuvels_ea:2014} for more details).
These stars apparently possess some degree of internal differential rotation, with a ratio $\Omega_g/\Omega_p$ in the range $2.5$--$20$, which increases with decreasing $\log g$, i.e., while the stars evolve away from the MS towards the red giant branch.

For the stars in the \citetalias{Mosser_ea:2012} sample, values of $\log g$ and $\Omega_g$ obtained from the ensemble asteroseismology technique (see their paper for details) are available.
This sample contains stars whose evolutionary stage ranges from the early red giant to the red clump. 
While the red giants (diamond symbols in Figure \ref{data}) are roughly in the same mass range as those of the \citetalias{Deheuvels_ea:2014} sample, the clump stars (triangle symbols) are much less homogeneous in mass, containing both low- and intermediate mass stars.
For this reason, only the red giants of the \citetalias{Mosser_ea:2012} sample will be considered here.

These two samples offer the possibility of a quantitative comparison between observed and modeled $\Omega_g/\Omega_p$ as a function of $\log g$, to constrain the efficiency of internal angular momentum transport through the poMS evolution.
As can be seen from Figure \ref{data}, after the TAMS a transition from the core spin-up of the subgiants to core spin-down among the red giants is observed.

In the following, we will compare the observed $\Omega_g$ and $\Omega_p$ with the average rotation rates over the radiative interior and the convective envelope, $\Omega_{\rm rad}$ and $\Omega_{\rm env}$ respectively, extracted from our models. 
It should be noted, however, that the $g$-mode cavity covers only part of the radiative interior, and therefore $\Omega_g$ does not exactly translate to $\Omega_{\rm rad}$.
We will neglect this effect from now on (see the discussion in section 6 of \citetalias{Deheuvels_ea:2014} for details).

\section{Modeling poMS rotational evolution of low-mass stars}
\label{models}

\subsection{The stellar evolution code}

The models discussed here were constructed using the YREC stellar evolution code (see \citealt{Demarque_ea:2008}, for a description of its non-rotational configuration, and \citealt{Endal_Sofia:1976,Pinsonneault_ea:1989}, for the treatment of rotation-related physics). 

We use the OPAL 2005 Equation of State \citep{Rogers_Nayfonov:2002}, and the OPAL Rosseland opacities \citep{Iglesias_Rogers:1996}, complemented by the \citet{Ferguson_ea:2005} opacities at low temperatures; the nuclear energy generation rates are those recommended by \citet{Adelberger_ea:2011}. 
The surface boundary conditions are based on the classical Eddington gray $T$--$\tau$ relationship.
Convection is described with the mixing length theory \citep{BV58}. 
We adopt the \citet{Grevesse_Noels:1993} value of the solar metallicity, $(Z/X)_\odot = 0.0245$.
The resulting solar-calibrated value of the mixing length parameter (the ratio of the mixing length over the pressure scale height) is $\alpha=1.832$, which is adopted throughout in our modeling.
To keep the number of parameters to a minimum, the effect of elements diffusion and convective core overshooting are ignored \citep[see][for a discussion of the effects of microphysics on the angular momentum transport]{Marques_ea:2013}.

In rotating models, the effect of rotation on the stellar structure, the angular momentum loss from the surface (if present), and the internal redistribution of angular momentum must also be taken into account.

In the treatment of the structural effects of rotation (i.e., increase in effective temperature, decrease in luminosity, increase of the MS lifetimes) we follow the standard YREC implementation.
These effects are quite small in low-mass stars (see, e.g., figures $4$ and $5$ of \citealt{Sills_ea:2000}).

For the wind braking we adopt the parametrization of \citet{Kawaler:1988}:
\begin{equation}
\label{kawaler}
\left(\frac{dJ}{dt}\right)_{\rm wind} = K_W \left(\frac{R_*/R_\odot}{M_*/M_\odot}\right)^{1/2}\, \Omega_*^3,
\end{equation}
where $R_*$, $M_*$, $\Omega_*$ are the radius, mass, and surface rotation rate of the star, and the overall scaling factor $K_W$ is an adjustable parameter.
The dependence on $\Omega_*^3$ of equation~\eqref{kawaler} results in an asymptotic rotational evolution that follows the empirically well-supported $P_{\rm rot} \propto t^{1/2}$ relation, where $P_{\rm rot}$ is the surface rotation period of the star and $t$ its age \citep{Skumanich:1972}. 
The second factor to the right hand side of equation~\eqref{kawaler}, containing $R_*$ and $M_*$, has been shown not to capture the full mass dependence of the wind braking phenomenon as observed in young open clusters \citep[see][for details]{Barnes_Kim:2010,Barnes:2010,Meibom_ea:2015,Lanzafame_Spada:2015}.
To compensate for the incorrect mass dependence in equation~\eqref{kawaler}, \cite{Chaboyer_ea:1995a,Chaboyer_ea:1995b} introduced a ``saturation phase", of mass-dependent duration (during which $\dot J_{\rm wind} \propto \Omega_*$). 
However, since for a star of mass $\gtrsim 1.1\, M_\odot$ the saturation phase only lasts for the first few Myr of the MS rotational evolution, to keep the number of parameters to a minimum we absorb both the saturation effect and the residual mass dependence of equation~\eqref{kawaler} within $K_W$, and retain it as the only freely adjustable parameter. 
In the following, to keep $K_W$ of the order of unity for convenience, we scale it over $1.13 \cdot 10^{47}$ g cm$^2$s, the value recommended by \citet{Kawaler:1988}.
 
We describe the radial transport of angular momentum in the stellar interiors as a diffusion process:
\begin{equation}
\label{amdiff}
\rho r^4 \frac{\partial \Omega}{\partial t} = \frac{\partial}{\partial r}\left[ \rho r^4 D\, \frac{\partial \Omega}{\partial r} \right],
\end{equation}
where $\Omega(r,t)$ is the angular velocity, $\rho$ is the density, and $D$ is the angular momentum diffusion coefficient.
Note that the values of $D$ adopted in the following are in any case representative of a turbulent process (i.e., much stronger than those resulting from just the molecular viscosity).
Solid-body rotation is enforced at all times within convective zones, as their typical mixing time scales are much faster than the processes discussed here.
The generation of latitudinal shear in convection zones does not affect their total angular momentum, and is therefore not relevant to our model.

In the standard version of YREC, the coefficient $D$ is calculated at each evolutionary time step by taking into account several hydrodynamical instabilities \citep{Pinsonneault_ea:1989}. 
Purely hydrodynamic instabilities, however, have already been shown not to be sufficiently effective at transporting angular momentum to explain the internal rotation of red giants \citep{Eggenberger_ea:2012,Ceillier_ea:2013,Marques_ea:2013}.
For this reason, here we follow a different approach.
We explore a simple, power law dependence of $D$ on the internal differential rotation, as measured by the ratio of the average angular velocity of the radiative interior, $\Omega_{\rm rad}$, over that of the convective envelope, $\Omega_{\rm env}$:
\begin{equation}
\label{dscaling}
D = D_0 \, \left( \frac{\Omega_{\rm rad}}{\Omega_{\rm env}} \right)^\alpha.
\end{equation}
In the equation above, $D_0$ and $\alpha$ are parameters to be determined that set the overall scale factor and the sensitivity of the dependence on the internal differential rotation, respectively.
A possible physical interpretation of such a dependence will be proposed in Section \ref{discussion}.

By determining the value of $\alpha$ that leads to the best agreement of the models with the data, we can obtain useful clues on the physical nature of the rotational coupling mechanisms in poMS stars.

\subsection{The solar benchmark}
\label{solcal}

In order to set baseline values of the constants $K_W$ and $D_0$, we apply the constant $D$ prescription, i.e., $\alpha=0$ in equation \eqref{dscaling}, to the MS rotational evolution of the Sun.

From the full solution for $\Omega(r,t)$ obtained for a $1\, M_\odot$ model, we extract the average rotation rate of the radiative and convective zones, $\Omega_{\rm rad}(t)$ and $\Omega_{\rm env}(t)$, as a function of time.
Simple, unweighted averages over the mass shells belonging to the radiative interior and to the convective envelope, respectively, are used.

The initial conditions are chosen such that the period at $1$ Myr is $8$ days. 
This choice roughly coincides with the median of the observed period distribution in the Orion Nebula Cluster \citep[ONC;][]{Rebull:2001}.
We adjust the values of $K_W$ and $D_0$ in order to satisfy the following constraints, which are meant to represent the present solar rotation \citep[e.g.,][]{Schou_ea:1998}:
\begin{equation*}
\Omega_{\rm env}(t_\odot) = 1.0 \, \Omega_\odot \ ;
\ \ \
\Omega_{\rm rad}(t_\odot)/\Omega_{\rm env}(t_\odot) = 1.02,
\end{equation*}
where $t_\odot = 4.57$ Gyr and $\Omega_\odot=2.78\cdot 10^{-6}$ s$^{-1}$.
The value adopted for $\Omega_{\rm rad}(t_\odot)/\Omega_{\rm env}(t_\odot)$ was estimated via numerical integration, using the fit of the solar rotation profile given by equations (14) and (15) of \citet{Roxburgh:2001}.
We thus obtain:
\begin{equation*}
K_{W,\odot} = 3.2; \ \ \ \ \ D_{\rm 0,\odot} = 2.5 \cdot 10^5\, {\rm cm}^2{\rm s}^{-1}. 
\end{equation*}
The parameter $K_W$ is mostly determined by the condition on $\Omega_{\rm env}(t_\odot)$, but it is also moderately sensitive to the behavior of internal differential rotation. 
For comparison, solid-body rotation is enforced at all times during the evolution if $D_0 \gtrsim 10^6$ cm$^2$ s$^{-1}$ \citep[see also][]{Denissenkov_ea:2010}, in which case our solar calibration gives $K_{W,\odot} = 2.9$.

Our solar-calibrated $D_{\rm 0,\odot}$ is in good agreement with previous studies \citep[e.g.][]{Ruediger_Kitchatinov:1996,Denissenkov_ea:2010,Spada_ea:2010}.
Another term of comparison is the value of $3\cdot 10^4$ cm$^2$ s$^{-1}$, which was found by \citet{Eggenberger_ea:2012} to reproduce the observed rotational splittings of KIC $8366239$.

\begin{figure}
\begin{center}
\includegraphics[width=0.49\textwidth]{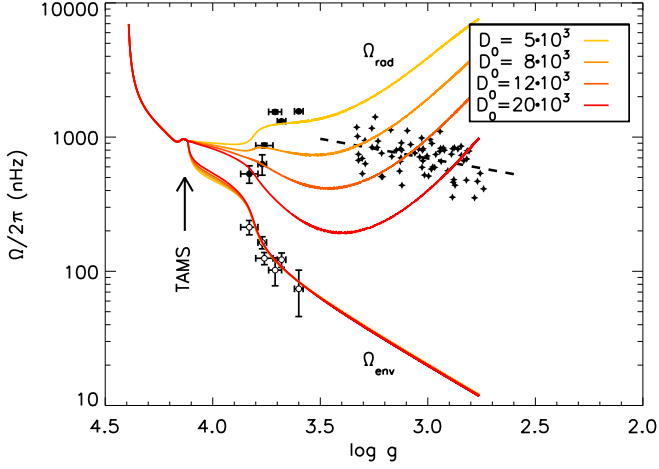}
\caption{PoMS rotational evolution of a $1.25\, M_\odot$ model with constant angular momentum diffusion coefficient, compared with the subgiant stars of \citetalias{Deheuvels_ea:2014} (open circles: $\Omega_p$; filled circles: $\Omega_g$)  and the red giants from \citetalias{Mosser_ea:2012} (diamonds). The broken line is a linear fit (in the variables $\log \Omega/2\pi$ vs. $\log g$) of the core rotation of the red giants. Values of $D_0$ in the legend are in cm$^2$s$^{-1}$.}
\label{constant_diffc}
\end{center}
\end{figure}

\begin{figure}
\begin{center}
\includegraphics[width=0.49\textwidth]{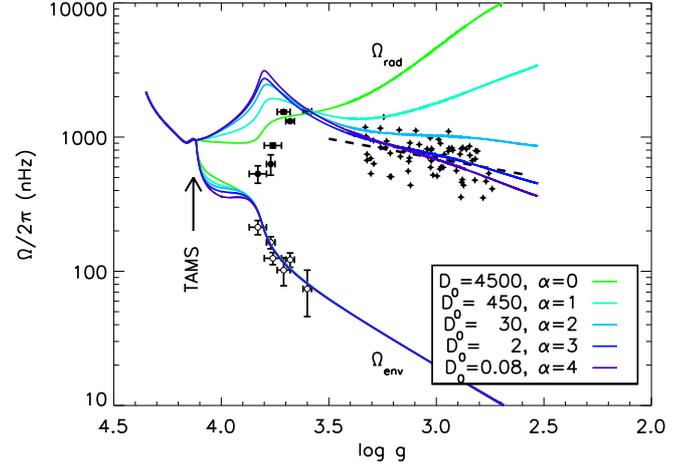}
\includegraphics[width=0.49\textwidth]{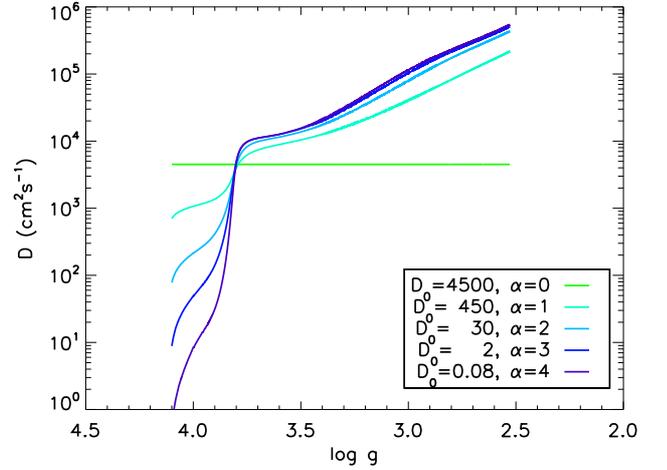}
\caption{PoMS rotational evolution models with angular momentum diffusion coefficient scaling with the ratio $\Omega_{\rm rad}/\Omega_{\rm env}$ according to equation \eqref{dscaling}. A solid-body interior rotation profile is enforced until the TAMS.
Upper panel: as in Figure \ref{constant_diffc}, but showing various non-zero $\alpha$ models as well. For each value of $\alpha$, the scaling parameter $D_0$ is calibrated to match the most evolved star in the \citetalias{Deheuvels_ea:2014} sample. 
Lower panel: evolution of $D$ as a function of $\log g$.
The values of $D_0$ shown in the legend are in cm$^2$s$^{-1}$.}
\label{varalpha_diffc}
\end{center}
\end{figure}

\section{Results}
\label{results}

We model the poMS rotational evolution of a solar composition, $1.25\, M_\odot$ star. 
This choice of parameters is assumed to be representative of both the \citetalias{Deheuvels_ea:2014} sample and of the red giants in the \citetalias{Mosser_ea:2012} sample.
Modelling the red clump stars of \citetalias{Mosser_ea:2012} is outside the aims of this paper; more massive stars, such as those in the sample of \citet{Deheuvels_ea:2015}, will be the subject of a subsequent investigation.

\subsection{Evolution of $\Omega_{\rm env}$ and calibration of $K_w$}

The rotational evolution is an initial value problem, requiring suitable initial conditions.
We evolve our $1.25\, M_\odot$ model starting from the early PMS (age $\lesssim 1$ Myr).
This choice has several advantages: since the initial model is fully convective, we can assume it has a solid-body rotation profile; moreover, due to the strong convergence properties of the $\Omega$ dependence in equation \eqref{kawaler}, the subsequent evolution is not too sensitive to the details of the initial conditions. 
Similarly to the solar benchmark model discussed in Section \ref{solcal}, the initial period is assigned so that the period at the age of $1$ Myr is $8$ days. 
This is still compatible with the observed rotation period distribution in ONC \citep{Rebull:2001}, at least within the level of accuracy with which its mass dependence is currently known. 
Furthermore, rigid rotation within the model is enforced until the end of the MS (i.e., until the central hydrogen abundance drops below $\approx 10^{-4}$). 
Indeed, theoretical models of stars of mass $\gtrsim 1\, M_\odot$ predict that they attain a quasi-rigid rotation state by the time they have reached their mature MS \citep[i.e., once they  are older than $\approx 1$ Gyr: see][]{Gallet_Bouvier:2013, Lanzafame_Spada:2015}. 
This is also observed in the Sun and in other solar-like stars \citep{Nielsen_ea:2014,Benomar_ea:2015}.
Our poMS rotational evolution modeling thus begins from a (realistic) state of negligible internal differential rotation. 
At the same time, the problem at hand has been effectively decoupled from the MS rotational evolution problem, where angular momentum redistribution is a highly debated, currently unsettled issue in itself \citep[e.g.][]{Charbonneau_MacGregor:1993,Ruediger_Kitchatinov:1996,Mathis_ea:2008, Garaud_Garaud:2008, Denissenkov_ea:2010}.

The parameter $K_W$ can be fixed by requiring that the evolution of $\Omega_{\rm env}(t)$ extracted from the models matches that of $\Omega_p$ of the subgiants in the $\log g$-$\Omega$ plane.
As can be seen from Figure~\ref{constant_diffc}, adjusting the single parameter $K_W$ produces a remarkably good fit of all the $\Omega_p$ in the \citetalias{Deheuvels_ea:2014} sample.
This implies that the surface spin-down of these subgiants is still in the Skumanich regime (i.e., $P \propto t^{1/2}$). 

\begin{figure}
\begin{center}
\includegraphics[width=0.49\textwidth]{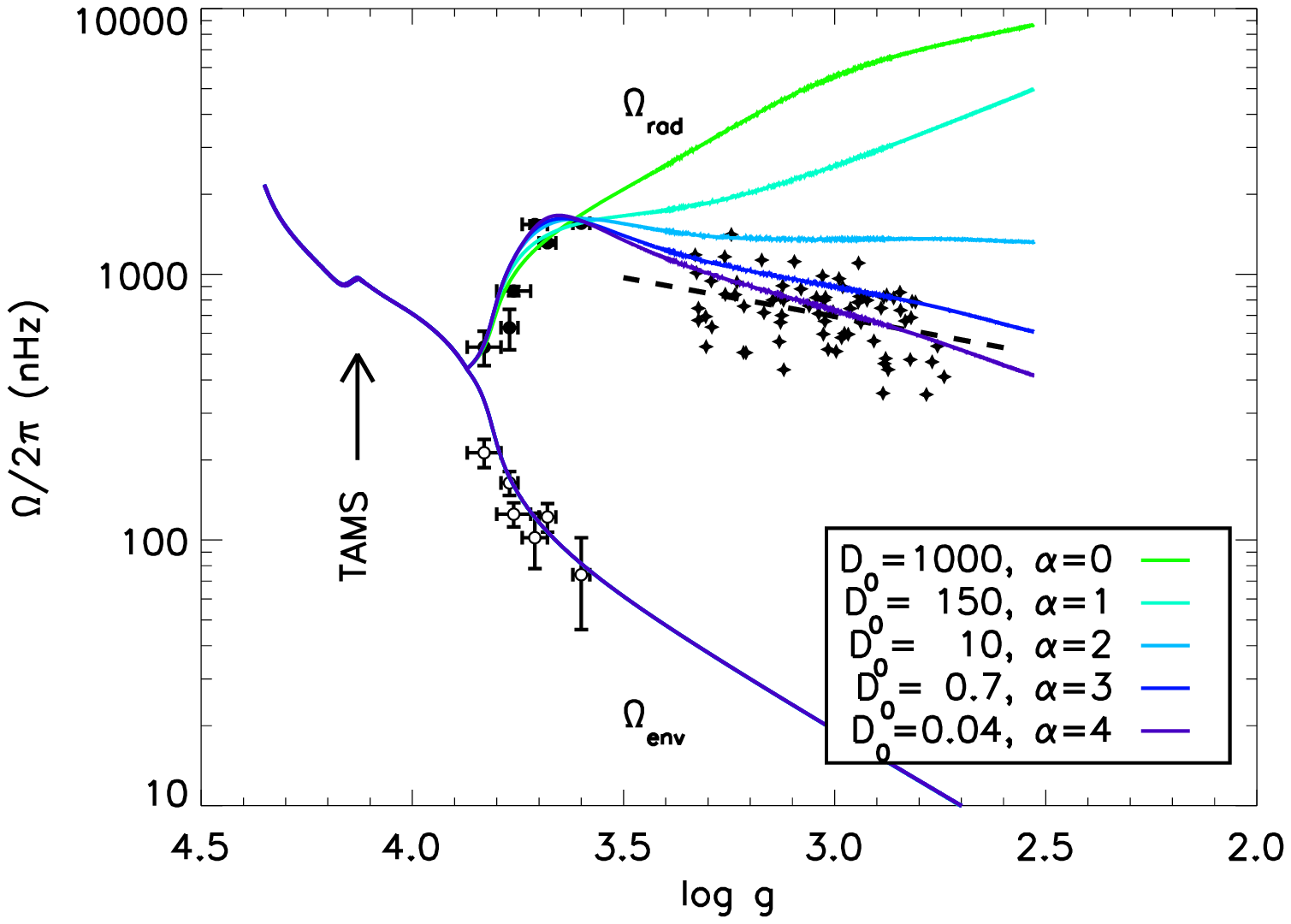}
\includegraphics[width=0.49\textwidth]{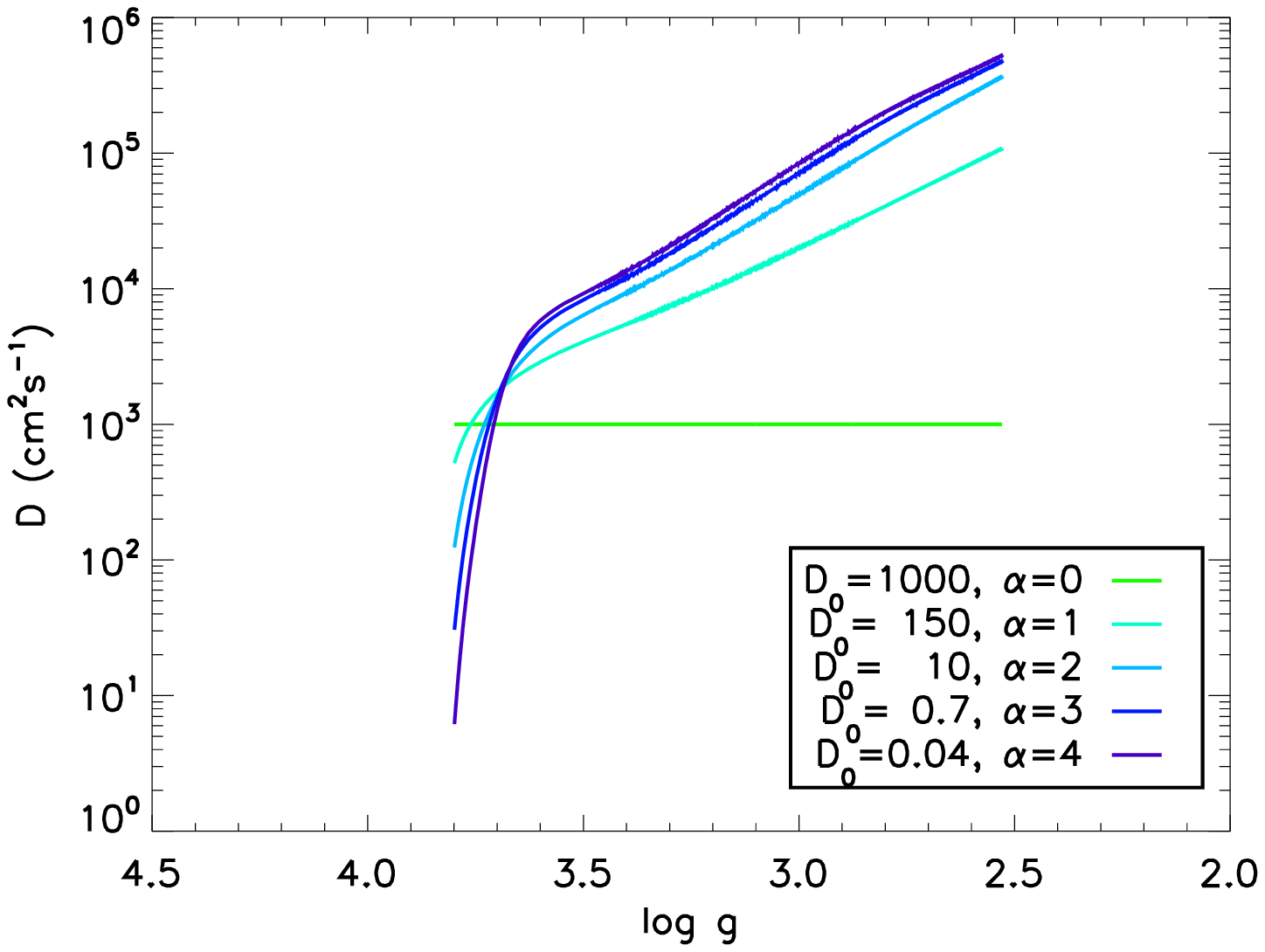}
\caption{  
As in Figure \ref{varalpha_diffc}, but showing models kept in a state of rigid rotation until approximately $1$ Gyr after the TAMS. 
Upper panel: poMS rotational evolution for various values of $\alpha$. For each $\alpha$, $D_0$ has been calibrated to obtain an optimal fit of both the subgiants and the red giants.
Lower panel: the evolution of the angular momentum diffusion coefficient according to equation \eqref{dscaling}.
}
\label{delay}
\end{center}
\end{figure}

The calibration of $K_W$ depends on the mass of the model, but is insensitive to the value of $D_0$, as can be seen from the overlap of the $\Omega_{\rm env}$ evolutions plotted in Figure \ref{constant_diffc}, or even to the prescription used for the angular momentum diffusion coefficient (compare Figure \ref{constant_diffc} with Figures \ref{varalpha_diffc} and \ref{delay}).
For $M_*=1.25\, M_\odot$ we obtain $K_W = 0.8$. 
This value of $K_W$ is kept fixed in all the following calculations.

\subsection{Constant angular momentum diffusion coefficient}
\label{constant}

We first discuss models with a diffusion coefficient independent of differential rotation (i.e., $\alpha=0$ in equation \ref{dscaling}).
The resulting poMS rotational evolution for several values of $D_0$ is shown in Figure~\ref{constant_diffc}.

From the Figure we can draw two main conclusions: first, although the overall behavior is not captured at all, values of $D_0 \approx 10^4$ cm$^2$s$^{-1}$ roughly match the order of magnitude of both the subgiants and the red giants core rotation (for comparison, the typical value of the molecular viscosity in the radiative zone is $2$--$20$ cm$^2$s$^{-1}$).
This is significantly smaller than what is required during the MS evolution of the Sun to match the helioseismic constraints (by a factor of $25$ according to our estimate of Section \ref{solcal}), but comparable to the value found by \citet{Eggenberger_ea:2012} for KIC $8366239$ ($3\cdot 10^4$ cm$^2$s$^{-1}$).
In other words, the poMS angular momentum transport is, overall, less efficient than during the MS \citep[see also the discussion in][]{Tayar_Pinsonneault:2013}. 

Secondly, as was already noted by \citet{Cantiello_ea:2014}, with a constant angular momentum diffusion efficiency it is impossible to satisfactorily reproduce both the subgiant core spin-up and the subsequent spin-down occurring during the red giant phase.
The small value of $D_0$ required to allow the development of internal differential rotation shortly after the TAMS results in a strong monotonic increase of the core rotation rate at later times, at odds with the trend observed in the \citetalias{Mosser_ea:2012} red giants.

\subsection{Angular momentum diffusion dependent on differential rotation}
\label{varalpha}

We now discuss the effect of a dependence of the angular momentum diffusion coefficient on the degree of internal differential rotation, in the form of equation \eqref{dscaling}.
The resulting poMS rotational evolutions for $\alpha=0,1,2,3,$ and $4$ are shown in the upper panel of Figure \ref{varalpha_diffc}.
At first, we do not attempt to fit the rotational evolution of the subgiants in detail, but rather the overall spin-up of the core during the subgiant phase, as represented by the most evolved star of the \citetalias{Deheuvels_ea:2014} sample, and the subsequent core spin-down.
This calibration fixes the value of $D_0$, given $\alpha$. 
A more satisfactory fit of all the stars in the \citetalias{Deheuvels_ea:2014} sample will be presented in Section \ref{prolonged}.
The lower panel of the Figure shows the corresponding evolution of $D$ according to equation \eqref{dscaling}.
Note that, as a result of the calibration of $D_0$ chosen, the diffusion coefficient has approximately the same value ($\approx 4.5\cdot 10^3$ cm$^2$s$^{-1}$) around $\log g=3.8$ in all the models.

The rotational evolution from the early red giant phase onwards ($\log g \lesssim 3.5$) is very sensitive to $\alpha$.
Qualitatively, the slope of the fit to the red giants trend (shown as a broken line in the Figure) is best reproduced with $\alpha\approx 3$.

\subsection{Prolonged post-TAMS solid-body rotation phase}
\label{prolonged}

Introducing a dependence of the angular momentum diffusion coefficient on the internal differential rotation according to equation \eqref{dscaling} allows the models to reproduce quite well the core spin-down trend on the red giant branch.
The agreement with the subgiant data is, however, not equally satisfactory.
From Figures \ref{constant_diffc} and \ref{varalpha_diffc} it is apparent that, as soon as the solid-body rotation constraint is relaxed, $\Omega_{\rm rad}$ and $\Omega_{\rm env}$ decouple very quickly, and the values of {\bf  $D_0$} required by the overall fit allow the ratio $\Omega_{\rm rad}/\Omega_{\rm env}$ to reach too large values in comparison with the youngest star in the \citetalias{Deheuvels_ea:2014} sample.

In the models shown so far, a solid-body rotation profile is artificially enforced until the TAMS. 
In this Section, we explore the possibility of a prolonged rigid rotation phase extending beyond the TAMS, of adjustable duration.
Indeed, as was discussed in the Introduction, there are good theoretical and observational reasons \citep{Gallet_Bouvier:2013,Lanzafame_Spada:2015,Nielsen_ea:2014,Benomar_ea:2015} to assume that stars of mass $\gtrsim 1\, M_\odot$ attain a rotationally coupled state on their mature main sequence. 
There are, however, no constraints on whether this state should cease immediately at the end of core hydrogen burning or last longer.

\begin{figure}
\begin{center}
\includegraphics[width=0.49\textwidth]{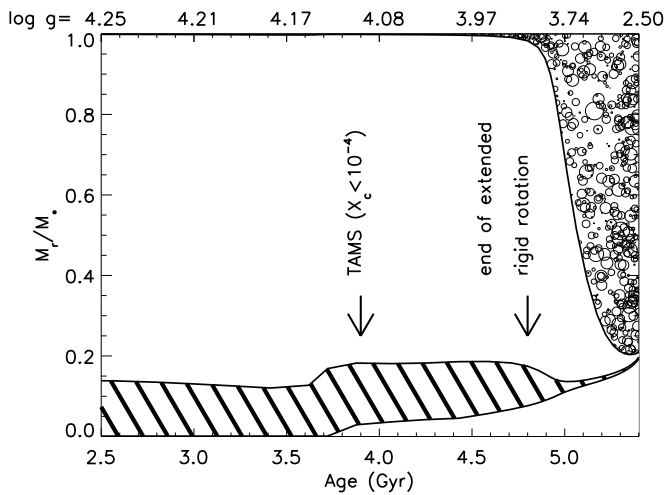}
\caption{Kippenhahn diagram (run of interior structure quantities as a function of mass coordinate and age) for a $1.25\, M_\odot$ model. The main region of hydrogen burning (i.e., where $\varepsilon_{\rm nucl}>10$ erg s$^{-1}$g$^{-1}$) is hatched; ``cloudy" areas indicate convection.}
\label{kipp}
\end{center}
\end{figure}

In Figure \ref{delay} we plot the results obtained with a prolonged rigid rotation phase. 
The best agreement with the \citetalias{Deheuvels_ea:2014} subgiants is found when enforcing solid-body rotation until the age of $\approx 4.8$ Gyr.
This roughly corresponds to the age at which shell hydrogen burning has become well established (see Figure \ref{kipp}).
For future reference, this occurs when the stellar radius has expanded to $\approx 2.16\, R_\odot$, or about $1$ Gyr after the end of core hydrogen burning.  
Both $\alpha=0$ and $\alpha=1,2,3,4$ models are shown in Figure \ref{delay}. 
This time, for each value of $\alpha$, $D_0$ has been calibrated to optimize the overall fit.
The lower panel of Figure \ref{delay} shows the corresponding evolution of the angular momentum diffusion coefficient $D$.
Note that the tuned values of $D_0$ reported in Figure \ref{delay} are smaller by a factor of $\approx 3$ than those in Figure \ref{varalpha_diffc}. 
This is due to the prolonged rigid rotation phase that results in a weaker differential rotation to be quenched afterwards.

A remarkable agreement with the data, from the subgiants all the way to the red giant branch, is achieved for $\alpha\approx 3$, $D_0\approx 1$ cm$^2$s$^{-1}$.

\section{Discussion}
\label{discussion}

The results of the previous section show that, during the poMS evolution of low-mass stars, the rotational coupling efficiency required to explain the observations is compatible with some turbulent process (i.e., it is enhanced compared to the molecular viscosity) and it is not constant in time.
The models including a power law dependence of the angular momentum diffusion coefficient on the ratio $\Omega_{\rm rad}/\Omega_{\rm env}$ (see equation \ref{dscaling}) can reproduce this behavior with $D_0\approx 1$ and $\alpha \approx 3$. 
As the lower panels of Figures \ref{varalpha_diffc} and \ref{delay} show, $D$ becomes essentially independent of $D_0$ and $\alpha$ during the red giant evolution for $\alpha \gtrsim 2$. 
This is a consequence of the differential rotation feedback introduced by equation \eqref{dscaling}, which allows the angular momentum transport efficiency to regulate itself and to reach a quasi-stationary state.

Various angular momentum transport processes in stellar interiors depend on the internal shear: for example, purely hydrodynamic instabilities, such as the dynamical and secular shear instability \citep[e.g.,][]{Zahn:1974}, or magnetohydrodynamic instabilities, such as the Tayler instability \citep{Tayler:1957,Tayler:1973} or the AMRI \citep{Ruediger_ea:2007,Ruediger_ea:2015}.

For the AMRI, in particular, a scaling of the turbulent viscosity in terms of dimensionless numbers has been established through direct numerical simulations performed in a Taylor-Couette cylindrical setup, taking into account the thermal stratification of the background fluid (Gellert et al., in preparation).
According to these simulations, the enhancement of the turbulent viscosity with respect to its molecular value, $\nu_T/\nu$, is given by:
\begin{equation}
\label{nut}
\frac{\nu_T}{\nu} \propto \sqrt{\frac{\rm Pm}{\rm Ra}} \; \frac{{\rm Re}}{\mu_\Omega^2},
\end{equation}
where ${\rm Pm} \equiv\nu/\eta$ is the Prandtl number, ${\rm Re} \equiv\Omega_i R_i (R_o-R_i) / \nu$ the Reynolds number, ${\rm Ra}\equiv {\cal Q}\, g\, (T_o-T_i)\, (R_o-R_i)^3 / \nu \chi$ the Rayleigh number, and $\mu_\Omega\equiv\Omega_o/\Omega_i$. 
In the previous definitions, $g$ is the acceleration of gravity, $\eta$ the magnetic diffusivity, $\chi$ the thermal conductivity, and $\cal Q$ the thermal expansion coefficient of the fluid; $R_i$, $T_i$, $\Omega_i$ and $R_o$, $T_o$, $\Omega_o$ are the radius, temperature, and angular velocity at the inner and outer boundary of the simulation domain, respectively.
The dependence of $\nu_T/\nu$ on rotation only is thus:
\begin{equation}
\frac{\nu_T}{\nu} \propto \frac{\Omega_i}{(\Omega_o/\Omega_i)^2} = \frac{\Omega_i^3}{\Omega_o^2}.
\end{equation}
Since, as was shown in Section \ref{results}, our best-fitting model requires $D \propto (\Omega_{\rm rad}/\Omega_{\rm env})^\alpha$ with $\alpha\approx 3$, and $D$ becomes almost independent of $\alpha$ on the red giant branch for $\alpha\gtrsim 2$ (see Figure \ref{delay}), we suggest the loose identification $\Omega_i\approx \Omega_{\rm rad}$ and $\Omega_o\approx\Omega_{\rm env}$. 

\begin{figure}
\begin{center}
\includegraphics[width=0.49\textwidth]{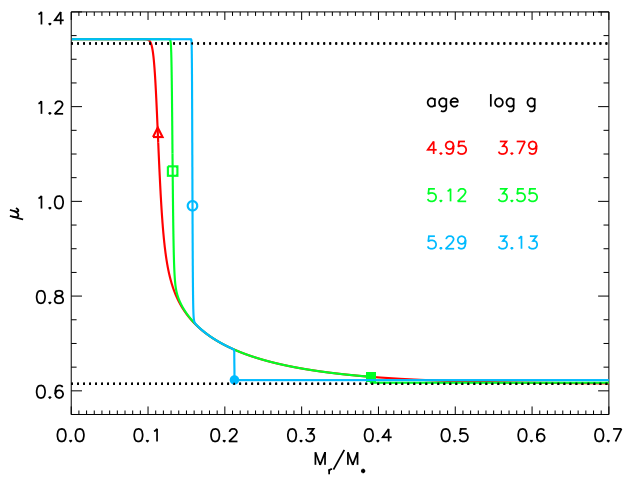}
\caption{
Mean molecular weight profile as a function of the fractional mass in the interior of $1.25 \, M_\odot$ models at three different evolutionary stages (cf. Figure \ref{data}): early post-TAMS (red, diamonds), subgiant (green, squares) and red giant (blue, circles). The approximate formula $\mu \approx (2X+\frac{3}{4}Y+\frac{1}{2}Z)^{-1}$, valid for a fully ionized gas, has been used. Ages in Gyr and $\log g$ are shown on the side. The filled and open symbols on each curve mark the He core outer boundary and the bottom of the outer convection zone, respectively. Dotted lines indicate the pure helium $\mu=4/3$ and pristine composition $\mu\approx 0.6$.
}
\label{mugrad}
\end{center}
\end{figure}

It must be emphasized, however, that, although the effectiveness of angular momentum transport by the AMRI in a chemically homogeneous fluid has been shown by \citet{Ruediger_ea:2015}, it is possible that the strong molecular weight gradient that develops in the core of a poMS star suppresses the instability.
To assess the importance of this effect, the profile of the mean molecular weight $\mu \approx (2X+\frac{3}{4}Y+\frac{1}{2}Z)^{-1}$ in the interior of our reference $1.25\, M_\odot$ model at three different evolutionary stages is plotted in Figure \ref{mugrad}.
The profiles shown in the Figure roughly correspond to models right after the TAMS, and in the subgiant and red giant stages.
As expected, the mean molecular weight ranges from $\mu\approx 1.33$ to $\mu\approx 0.6$ within the He core and in the convection zone, respectively. 
The strongest gradient is found at the transition between the pure helium core and the rest of the star.
Clearly, more extensive numerical and modeling work, taking into account the effect of $\mu$ gradients directly, is required to put the identification suggested above between our heuristic scaling \eqref{dscaling} and equation \eqref{nut} on firmer ground.

A satisfactory agreement of our models with both the subgiants in the \citetalias{Deheuvels_ea:2014} sample and the red giants in the \citetalias{Mosser_ea:2012} sample can only be obtained by assuming that the solid-body rotation regime achieved during the mature MS continues to hold until the nuclear energy source of the model has shifted from core to shell hydrogen burning. 
This occurs about $1$ Gyr after the TAMS in our $1.25\, M_\odot$ model.
Since there are currently no observational constraints on this issue, we can only propose some theoretical arguments.
The two leading explanations for the coupling in the interior of low-mass MS stars are magnetic fields \citep[e.g.,][]{Ruediger_Kitchatinov:1996} and internal gravity waves \citep[e.g.,][]{Mathis_ea:2008}.
In the case of the former, it is plausible that the action of magnetic fields continues beyond the end of the hydrogen burning phase, and becomes ineffective only when the star has undergone significant structural changes evolving towards the red giant branch.
For the latter, \citet{Fuller_ea:2014} have shown that internal gravity waves can affect internal rotation on a short ($\approx 100$ Myr) time scale on the MS, but that this time scale progressively increases, eventually leading to a decoupling between the stellar core and envelope.
They estimated that, for low-mass stars ($M\lesssim 1.5\, M_\odot$), this occurs when the stellar radius has increased to about $1.75$ times the MS radius. 
This is in very good agreement with our prolonged solid-body rotation scenario discussed in Section \ref{prolonged}, where the strong coupling regime was assumed to last until $R_*\approx 2.16\, R_\odot$ (note that our $1.25\, M_\odot$ model has a MS radius of about $1.2\, R_\odot$).
Either way, we could speculate that differential rotation begins to develop at some point after the TAMS, during the subgiant/early red giant phase, until angular momentum redistribution is taken over by some other, dominant process during the poMS.

Finally, we note that the overshooting at the bottom of the surface convection zone, which has been ignored in our calculations, could have a significant impact on our results if it can bring the bottom of the outer convection zone close enough to the hydrogen-burning shell, bridging the gap of the region where most of the shear develops{\footnote{Note that turbulent angular momentum transport can occur independently of chemical element mixing, see, e.g., \citet{VK83}.}.
As a crude estimate, we find that on a typical evolution, and considering an overshooting of $0.20$ pressure scale heights at the bottom of the outer convection zone, the overshooting layer is less than $\approx 20\%$ of the radial extension of the shear region.
This is not enough to change the qualitative picture of the evolution from the subgiant through the red giant stages.

\section{Conclusions}
\label{conclusions}

We have discussed poMS rotational evolution models for a solar composition, $1.25\, M_\odot$ star, representative of the low-mass component ($1.0 \lesssim M/M_\odot \lesssim 1.5$) of the asteroseismic analyses of \citet{Deheuvels_ea:2014} and \citet{Mosser_ea:2012}.

The models include a standard parametrization of the braking of the stellar surface due to the magnetized stellar winds. 
Angular momentum transport in the interior is treated as a diffusion process, implementing a simple formulation for a dependence of the diffusion coefficient on the internal differential rotation.

Our main conclusions are the following:
\begin{enumerate}
\item Angular momentum transport in the early poMS is less efficient, by more than one order of magnitude, than that required for the PMS and MS evolution of the Sun to match the helioseismic constraints;
\item Assuming an angular momentum diffusion coefficient constant in time results in a monotonic spin-up of the stellar core from the subgiant phase onwards, which is incompatible with the available observational constraints;
\item 
An angular momentum diffusion coefficient dependent on the internal shear can establish a quasi-stationary core spin-down regime during the red giant phase, and lead to a rotational evolution in agreement with the observations;
\item Full agreement between models and data from the TAMS all the way to giant branch evolution (before the red clump) can be achieved assuming that stars remain in a rigid rotation state until the shell hydrogen burning phase ($\approx 1$ Gyr after the TAMS for a $1.25\, M_\odot$ star). 
\end{enumerate}

\acknowledgements{
FS acknowledges support from the Leibniz Institute for Astrophysics Potsdam (AIP) through the Karl Schwarzschild Postdoctoral Fellowship.
MG would like to acknowledge support by the Helmholtz Alliance LIMTECH.}


 {}




\end{document}